\documentclass{article}
\usepackage{spconf,amsmath,graphicx}
\usepackage{todonotes}
\usepackage{multirow}
\usepackage{tabularx}
\usepackage{verbatim}
\usepackage{hyperref}
\usepackage{subfig}

\title{Semi-overcomplete convolutional auto-encoder embedding as shape priors for deep vessel segmentation}

\name{A. Sadikine $^{\bullet, \star}$
\qquad \hspace{-0.7cm} B. Badic $^{\bullet, \flat}$
\qquad \hspace{-0.7cm} J.-P. Tasu $^{\bullet, \circ}$
\qquad \hspace{-0.7cm} V. Noblet $^{\sharp}$
\qquad \hspace{-0.7cm} D. Visvikis $^{\bullet}$
\qquad \hspace{-0.7cm} P.-H. Conze $^{\bullet, \diamond}$
\thanks{This work was partially funded by LaBeX CAMI (grant ANR-11-LABX-0004) and France Life Imaging (grant ANR-11-INBS-0006).}}

\address{$^{\bullet}$ LaTIM UMR 1101, Inserm, Brest, France 
$^{\star}$ University of Western Brittany, Brest, France \\
$^{\flat}$ University Hospital of Brest, Brest, France
$^{\circ}$ University Hospital of Poitiers, Poitiers, France \\
$^{\sharp}$ ICube UMR 7357, CNRS, Strasbourg, France
$^{\diamond}$ IMT Atlantique, Brest, France \\
}

\begin{document}

\maketitle

\begin{abstract}
The extraction of blood vessels has recently experienced a widespread interest in medical image analysis. Automatic vessel segmentation is highly desirable to guide clinicians in computer-assisted diagnosis, therapy or surgical planning. Despite a good ability to extract large anatomical structures, the capacity of U-Net inspired architectures to automatically delineate vascular systems remains a major issue, especially given the scarcity of existing datasets. In this paper, we present a novel approach that integrates into deep segmentation shape priors from a Semi-Overcomplete Convolutional Auto-Encoder (S-OCAE) embedding. Compared to standard Convolutional Auto-Encoders (CAE), it exploits an over-complete branch that projects data onto higher dimensions to better characterize tiny structures. Experiments on retinal and liver vessel extraction, respectively performed on publicly-available DRIVE and 3D-IRCADb datasets, highlight the effectiveness of our method compared to U-Net trained without and with shape priors from a traditional CAE.
\end{abstract}

\begin{keywords}
vascular segmentation, shape priors, overcomplete representations, tubular structures
\end{keywords}

\begin{figure}[!t]
    \centering
    \includegraphics[scale=0.6]{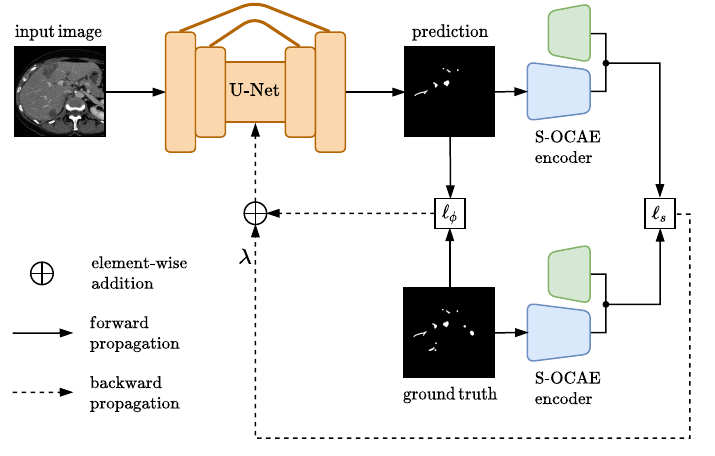}
    \caption{Proposed pipeline. U-Net parameters are estimated by penalizing a segmentation loss $\ell_{\phi}$ as well as a regularization term $\ell_{s}$ dealing with the similarity between projections of prediction and ground truth in a learned S-OCAE latent space.} \vspace{-0.2cm}
\end{figure}

\section{Introduction}
\label{sec:intro}

Analyzing vessels from medical images is an essential task for diagnosis, therapy or surgery planning purposes \cite{abramoff2010retinal, sahani2004preoperative}. Since manual segmentation is a time-consuming and complex process, there remains a great demand for an automated approach that could effectively capture vascular structure contours while managing low contrast, high level of noise,  intensity variations, multi-scale geometry and complex bifurcated topology \cite{l2017recent}. Over the past few years, deep learning (DL) has given a huge boost in semantic segmentation methodologies with remarkable performance. In the medical field, the U-Net convolutional architecture \cite{ronneberger2015unet} has gained great popularity for various clinical applications. However, despite a good ability to extract large structures such as organs \cite{kavur2021chaos,conze2021abdominal}, its capacity to delineate blood vessels remains a major challenge.

In the context of retinal vessel segmentation from 2D color fundus images, several DL methodologies has been developed including \cite{shin2019deep} which takes into account the graphical structure of vessel shape in addition to local vessel appearance information, and \cite{karaali2021dr} which exploits cascaded sub-networks to deal with tiny retinal blood vessel branches. Heavy data augmentation has been alternatively proposed to alleviate dataset limitations \cite{uysal2021exploring}. Among existing works on 3D vessel extraction, early attempts introduced three-plane segmentation approaches to reach 3D delineation results from 2D patch-based techniques \cite{kitrungrotsakul2019vesselnet, oda2019abdominal}. To get rid of intensity changes, VesselNet integrated vessel probability maps  as inputs to ease the extraction of tubular structures. \cite{kitrungrotsakul2019vesselnet}. Data imbalance issues was handled through an area imbalance reduced training patch generation \cite{oda2019abdominal}. Nevertheless, none of these approaches exploit geometric constraints to make fine vascular branches being detected and connected from main vessels.

Significant performance has been achieved through the incorporation of anatomical shape priors into medical image segmentation networks. In cardiac image analysis, Oktay et al. proposed to add a penalty term to the global loss function dealing with the Euclidean distance between the projection of both predicted and ground truth segmentation masks in a convolutional auto-encoder (CAE) latent space \cite{oktay2017anatomically}. This aims at guiding the segmentation model to follow the global anatomy of the target. To go further, shape priors and adversarial learning were combined as regularizers to encourage the network to provide realistic shapes for bone segmentation \cite{boutillon2020combining,boutillon2021multi}. To our knowledge, no study on the association of segmentation model and shape priors has been proposed in the literature for blood vessel delineation. In the meantime, several approaches have recently focused on the exploration of more sophisticated deep architectures than the standard U-Net \cite{ronneberger2015unet} via residual connections \cite{yu2019liver, karaali2021dr} or overcomplete representations \cite{valanarasu2020kiu}. Overcomplete architectures  \cite{valanarasu2020kiu, valanarasu2021overcomplete} have appeared with the goal of projecting data onto higher dimensions to constrain the receptive field to be small and therefore capture finer low-level features details. The ability of over-complete architectures to encode small anatomical structures appears of high interest for vessel extraction purposes. 

Since modeling small vascular structures with standard CAE  \cite{oktay2017anatomically,boutillon2020combining,boutillon2021multi} does not guarantee an efficient shape representation in latent space, we propose to integrate into the segmentation framework a new Semi-Overcomplete Convolutional Auto-Encoder (S-OCAE) with a multi-path encoder leveraging both non-linear under and overcomplete representations of the multi-scale vascular tree geometry, from tiny to large vessels. The effectiveness of the resulting semi-overcomplete shape priors is illustrated for two clinical applications dealing with retinal and liver vessel extraction.

\section{Methods}
\label{sec:methods}

Let us denote $\pmb{x}$ a greyscale image and $\pmb{y}$ its corresponding ground truth segmentation mask. Supervised segmentation with DL consists of approximating a mapping function $\phi:\pmb{x}\rightarrow \phi(\pmb{x})=\pmb{\hat{y}}$ from $q$ training samples $\{\pmb{x}_{i},\pmb{y}_{i}\}_{i<q}$ by optimizing a loss function $\ell_{\phi}(\pmb{y},\pmb{\hat{y}})$ through a stochastic optimizer. We define $\phi$ as a U-Net shaped segmentation model, a convolutional encoder-decoder made of both contracting and expansive paths with lateral skip-connections \cite{ronneberger2015unet}. A regularization strategy arising from shape prior information \cite{oktay2017anatomically} can particularly help in the delineation of vessels whose multi-scale shape characteristics are very complex to manage.

\subsection{Standard undercomplete shape priors}

To deal with shape priors, one can exploit a compact representation of the anatomy coming from ground truth segmentation masks using a CAE \cite{oktay2017anatomically} consisting of an encoder $E$ (parameterized by $\pmb{\Theta}_{E}$) and a decoder $D$. The encoder transforms a ground truth segmentation mask $\pmb{y}$ through a cascade of convolutional, batch normalization (BN), non-linearity (e.g. ReLU) and max-pooling layers to a compressed representation $E(\pmb{y};\pmb{\Theta}_{E})=\pmb{z}$ referred to as latent code. Conversely, the decoder consists of decoding the information from the latent code $\pmb{z}$ through series of transposed and normal convolutions, BN and non-linearity to reconstruct the input $\pmb{\Tilde{y}}=D(\pmb{z})=D\circ E(\pmb{y})$. To train the CAE, we encourage it to accurately reconstruct its input segmentation masks by optimizing:  \vspace{-0.15cm}

\begin{equation}\label{mse}
    \ell_{CAE}(\pmb{y},\pmb{\Tilde{y}})\propto \sum_{i=0}^{q-1}{\|{\pmb{y}_i-\pmb{\Tilde{y}}_i}\|_{2}^{2}}
\end{equation} \vspace{-0.05cm}

Let $F_{E}^l$ and $F_{D}^{l\sp{\prime}}$ be symmetrical hidden layer outputs at depths $l$ and $l\sp{\prime}$ from $E$ and $D$, defined respectively as $\pmb{y}^l= F_{E}^l(\pmb{y})$ and $\pmb{z}^{l\sp{\prime}}\hspace{-0.075cm}= F_{D}^{l\sp{\prime}}(\pmb{z})$, with $F_{E}^l=E^{l} \circ\cdots\circ E^{1}$ and $F_{D}^{l\sp{\prime}}=D^{l\sp{\prime}}\circ\cdots\circ D^{1}$. The spatial dimension of $\pmb{y}^l$ and $\pmb{z}^{l\sp{\prime}}$ is far below the dimensionality of the input. For this reason, the CAE can be defined as an undercomplete auto-encoder. The first few layers $F_{E}^l$ aims at capturing low-level features and as $l$ increases, i.e. as we go deeper in the network, the model extracts more high-level features due to down-sampling operations which lead to larger receptive fields (RF). 

Once the standard CAE has been trained, shape priors can be incorporated into the segmentation pipeline by adding to the segmentation loss $\ell_{\phi}$ a shape regularization term $\ell_{s}$: \vspace{-0.25cm}

\begin{equation}\label{regterm}
    \ell_{s}(\pmb{y},\pmb{\hat{y}}) \propto \sum_{y}{( 1-\cos({E(\pmb{y};\pmb{\Theta}_{E}), E(\pmb{\hat{y}};\pmb{\Theta}_{E})}))}
\end{equation} \vspace{-0.25cm}

\noindent which measures the cosine distance between predicted outputs and ground truth masks in low dimensional space. The total loss is then expressed as follows:

\begin{equation}\label{globalloss}
   \mathcal{L} = \ell_{\phi}(\pmb{y},\phi(\pmb{x};\pmb{\Theta}_{\phi})) + \lambda\ell_{s}(\pmb{y},\phi(\pmb{x};\pmb{\Theta}_{\phi}))
\end{equation} \vspace{-0.2cm}

\noindent where $\pmb{\Theta}_{\phi}$ denotes all trainable parameters of $\phi$, $\lambda$ is a hyper-parameter empirically set weight. Incorporating $\ell_{s}$ into Eq.\ref{globalloss} constrains $\phi(\pmb{x};\pmb{\Theta}_{\phi})$ to capture more shape features from tubular targets. This strengthens plausible shape delineations in predicted masks and reduces false-positive detections. Optimal parameters $\pmb{\hat{\Theta}}_{\phi}$ are obtained by minimizing Eq.\ref{globalloss}:

\begin{equation}
   \pmb{\hat{\Theta}}_{\phi}=\arg \min_{\pmb{\Theta}_{\phi}} \mathcal{L}
   \label{optimalparam}
\end{equation} \vspace{-0.2cm}

\begin{figure}[t]
    \centering
    \includegraphics[scale=0.6]{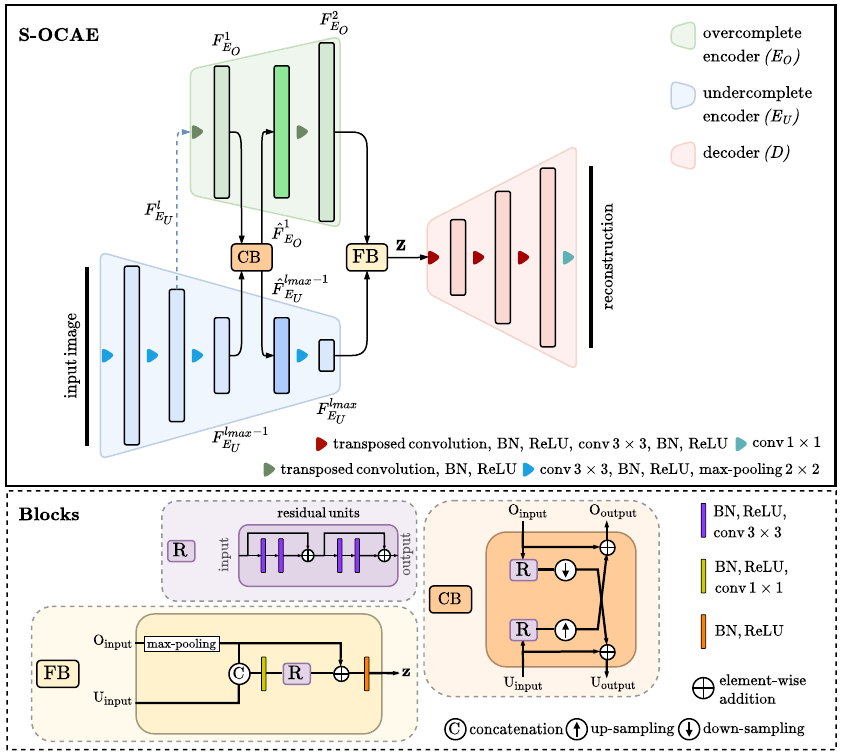}
    \caption{Proposed S-OCAE network whose multi-path encoder is made of undercomplete and overcomplete branches and includes  communication (CB), fusion (FB) and residual blocks.}
    \label{fig:fig2}
\end{figure} \vspace{-0.25cm}

\subsection{Extension to semi-overcomplete shape priors}

Compared to standard undercomplete CAE, an overcomplete CAE can be obtained by replacing max-pooling by up-sampling layers and vice-versa. In this scenario, intermediate layers $\pmb{y}^l$ and $\pmb{z}^{l\sp{\prime}}$ are projected to a higher dimensionality than the input. This gives to the model a better flexibility to capture and encode low-level features since RF are constricted.

In the same spirit of \cite{valanarasu2021overcomplete}, we propose to extend traditional CAE by exploiting a multi-path encoder composed of an undercomplete branch $E_U$ and an overcomplete branch $E_O$. Since an overcomplete encoder would be too expensive in terms of memory, we proposed a semi-overcomplete branch which takes as input hidden layer $F_{EU}^l$ of $E_U$ instead of the input image (Fig.\ref{fig:fig2}). $E_U$ allows to aggregate global information due to the expansion of RF in deeper $F_{E_U}^l$ layers but is not able to accurately encode small or tiny structures. To alleviate this issue, the branch $E_O$ made of two layers $F_{E_O}^{1}$ and $F_{E_O}^{2}$ is added at $F_{E_U}^{l_{max}-1}$ and $F_{E_U}^{l_{max}}$ layer levels to restrict the RF. The resulting multi-path encoder therefore manages both under and overcomplete representations.

A specific block named communication block (CB) was designed to parallely communicate residuals features between $F_{E_U}^{l_{max}-1}$ and $F_{E_O}^{1}$, with identity skip connection, by adding to them residual features $[R_{J}]^{\downarrow \frac{1}{n}}$ and $[R_{J}]^{\uparrow n}$ respectively (Fig.\ref{fig:fig2}). This results in new feature maps $\hat{F}_{E_U}^{l_{max}-1}$ and $\hat{F}_{E_O}^{1}$ to be forwarded in the network: \vspace{-0.3cm}

\begin{equation}\label{CB}
    CB \overset{\Delta}{=}
    \begin{cases}
        \hat{F}_{E_U}^{l_{max}-1}=F_{E_U}^{l_{max}-1}\oplus[R_{J}]^{\downarrow \tfrac{1}{n}}\, &\hspace{-0.2cm}\textrm{s.t.}\,R_{0}\hspace{-0.1cm}=\hspace{-0.1cm}F_{E_O}^{1}\hspace{-0.3cm}\\
        \hat{F}_{E_O}^{1}\hspace{0.6cm}= F_{E_O}^{1}\hspace{0.6cm}\oplus[R_{J}]^{\uparrow n}\, &\hspace{-0.2cm}\textrm{s.t.}\,R_{0}\hspace{-0.1cm}=\hspace{-0.1cm}F_{E_U}^{l_{max}-1}\hspace{-0.3cm}
    \end{cases}
\end{equation}

\noindent where $\oplus$ is the element-wise addition,  $[\cdot]^{\downarrow \frac{1}{n}}$ and $[\cdot]^{\uparrow n}$ bilinear down-sampling and up-sampling operations with factors $\frac{1}{n}$ and $n$. $R_{J}=R_{J-1}\oplus\sum_{i=0}^{J}\mathcal{F} (R_{i},\mathit{\pmb{w}}_i)$ is a forward recursion  to compute a serie of $J$ full pre-activation (i.e. BN and ReLU come before convolution layers) residual units \cite{he2016identity} where $\mathcal{F}$ is a residual function with trainable parameters $\mathit{\pmb{w}}_i$ defined by a cascade of BN, ReLU and convolution layers.

To combine the outputs from undercomplete and overcomplete encoding branches, respectively $F_{E_U}^{l_{max}}$ and $F_{E_O}^{2}$, a fusion block (FB) was introduced at the bottleneck of the network (Fig.\ref{fig:fig2}) to compute a latent code $\pmb{z}$. FB projects $F_{E_O}^{2}$ to a lower dimension through max-pooling $\mathcal{M}_{s}^{c}(\cdot)$, with stride $s\hspace{-0.1cm}=\hspace{-0.1cm}16$ and pooling coefficient $c\hspace{-0.1cm}=\hspace{-0.1cm}2$, to enable the concatenation with $F_{E_U}^{l_{max}}$: $\pmb{z}^{\prime}=\texttt{concat}[F_{E_U}^{l_{max}},\mathcal{M}_{s}^{c}(F_{E_O}^{2})]$.  Then, a ${1\times1}$ convolution operation is applied to $\pmb{z}^\prime$ with full pre-activation to get $\pmb{z}^{\prime\prime}$ with the same number of feature maps as $F_{E_U}^{l_{max}}$. $R_{J}$ is finally applied (with $R_{0}=\pmb{z}^{\prime\prime}$) to obtain the residual feature $\pmb{z}^{\prime\prime\prime}$. Finally, full post-activation was applied to  $\pmb{z}^{\prime\prime\prime}\oplus\mathcal{M}_{s}^{c}(F_{E_O}^{2})$ to reach the final latent code, considered as a non-linear representation of the anatomical shape for a given input. In practice, CB and FB  blocks employed $J=2$ (Fig.\ref{fig:fig2}) which seemed optimal during network design.

\section{Experiments}
\label{sec:experiment}
\begin{table*}[!ht]
\centering
\caption{Quantitative comparisons of U-Net+S-OCAE (without and with CB) with U-Net \cite{ronneberger2015unet} and U-Net+CAE \cite{oktay2017anatomically} using Dice (\texttt{DSC}), absolute volume diff. (\texttt{AVD}), average symmetric surface (\texttt{ASSD}) and Hausdorff (\texttt{HD}) distances. Best results in bold.} \vspace{-0.1cm}
\resizebox{510pt}{!}{
\begin{tabular}{@{} p{2.5cm} c*{8}{c} @{} c}
&\multicolumn{4}{c}{\textbf{DRIVE} \cite{staal2004ridge}} & & \multicolumn{4}{c}{\textbf{3D-IRCADb} \cite{soler20103d}}\\[1mm]
\cline{2-5} \cline{7-10}
\noalign{\smallskip}
&\shortstack{\texttt{DSC}$\uparrow$ \\ score ($\%$)} &\shortstack{\texttt{AVD}$\downarrow$\\ dist. (mm)} &\shortstack{\texttt{ASSD}$\downarrow$\\ dist. (mm)}    &\shortstack{\texttt{HD}$\downarrow$\\ dist. (mm)} &&\shortstack{\texttt{DSC}$\uparrow$ \\ score ($\%$)} &\shortstack{\texttt{AVD}$\downarrow$\\ dist. (mm)} &\shortstack{\texttt{ASSD}$\downarrow$\\ dist. (mm)}& \shortstack{\texttt{HD}$\downarrow$\\ dist. (mm)}\\\hline
U-Net \cite{ronneberger2015unet}                 			&$75.14\pm1.68$     &$0.38\pm0.13$    &$0.71\pm0.24$  &$32.90\pm1.85$&&$54.81\pm1.87$    &$0.62\pm0.21$    &$4.25\pm0.42$  &$67.60\pm4.74$\\
U-Net+CAE \cite{oktay2017anatomically}            &$79.25\pm0.48$  &$0.10\pm0.05$      &$\mathbf{0.54}\pm0.11$      &$31.82\pm2.78$  &&$\mathbf{59.26}\pm1.33$      &$0.33\pm0.14$    &$3.95\pm0.21$  &$54.48\pm5.38$\\
Ours (w/o CB)   &$77.75\pm0.66$  &$0.13\pm0.05$      &$1.22\pm0.19$      &$30.08\pm2.30$  &&$58.65\pm1.09$      &$0.37\pm0.12$    &$\mathbf{3.79}\pm0.31$  &$54.85\pm8.86$\\
\textbf{Ours (w/ CB)}   &$\mathbf{79.44}\pm0.65$  &$\mathbf{0.08}\pm0.02$      &$\mathbf{0.54}\pm0.15$      &$\mathbf{28.74}\pm0.58$  &&$59.14\pm1.08$      &$\mathbf{0.31}\pm0.15$    &$3.84\pm0.30$  &$\mathbf{50.45}\pm8.99$\\\hline
\end{tabular}
}
\label{Tab:table} \vspace{-0.2cm}
\end{table*}

\subsection{Imaging datasets}

We evaluate the proposed method on two public datasets: 3D-IRCADb and DRIVE. 3D-IRCADb\footnotemark[1] \cite{soler20103d} is an abdominal dataset that contains CT scans with ground truth labels from 20 different patients (10 women, 10 men) with liver tumors in $75\%$ of cases. We resized all axial slices to $256\times256$ pixels after extracting a liver bounding box in each CT scan. DRIVE\footnotemark[2] \cite{staal2004ridge} is a retinal blood vessel dataset that contains 20 RGB images (including signs of mild early diabetic retinopathy) with associated ground truth. Color fundus photographs were resized from $584\times565$ to $512\times512$ pixels.

\footnotetext[1]{\url{https://www.ircad.fr/research/3d-ircadb-01/}}
\footnotetext[2]{\url{https://drive.grand-challenge.org/}}

\subsection{Implementation details}

In the first stage, S-OCAE was trained using mean squared error loss (Eq.\ref{mse}) with Adam optimizer. The learning rate, batch size and number of epochs were set to $0.001$ ($0.0005$), $4$ ($32$) and $1000$ ($100$) for DRIVE (3D-IRCADb). In the second stage, the segmentation model was trained by optimizing Eq.\ref{globalloss} with weighted binary cross-entropy for $\ell_{\phi}$. The learning rate, batch size and number of epochs were set to $0.001$ ($0.0001$), $4$ ($16$) and $200$ ($100$) for DRIVE (3D-IRCADb). Optimal $\lambda$ was empirically set to $40$ ($60$) for DRIVE (3D-IRCADb). Data augmentation was applied during training for both networks: rotation, shearing and translation with additional flip, blurring and Gaussian noise for DRIVE. We used 5-fold cross-validation to validate the results. Both S-OCAE and segmentation networks were implemented in PyTorch and respectively trained on a Nvidia-RTX 2080Ti and A6000Ti GPU.

\subsection{Evaluation of predicted segmentation}

In order to evaluate the performance of our model (U-Net+S-OCAE) against existing approaches (U-Net \cite{ronneberger2015unet}, U-Net+CAE \cite{oktay2017anatomically}), we compared ground truth \(\mathit{GT}\) and prediction \(\mathit{P}\), defined respectively by the surface \(\mathit{S_{GT}}\) and \(\mathit{S_{P}}\), through the following metrics: Dice coefficient (\(\frac{2|\mathit{GT}\cap \mathit{P}|}{|\mathit{GT}|+|\mathit{P}|}\)), absolute volume difference (\(\frac{||\mathit{GT}|-|\mathit{P}||}{|\mathit{GT}|}\)), average symmetric surface distance \(\mathit{ASSD(\mathit{GT}, \mathit{P})}=\frac{1}{|\mathit{S_{GT}}|+|\mathit{S_{P}}|}(\sum_{\mathit{s} \in \mathit{S_{GT}}}\mathit{d(\mathit{s}, \mathit{S_{P}})}+\sum_{\mathit{s} \in \mathit{S_{P}}}\mathit{d(\mathit{s}, \mathit{S_{GT}})})\) where \(\textstyle{\mathit{d(\mathit{s}, \mathit{S_{k}})}= min_{\mathit{s_k} \in \mathit{S_{k}}}\,\|\mathit{s}-\mathit{s_{k}}\|}\) and  \(\textstyle{\|\cdot\|}\) the Euclidean distance. The Hausdorff distance was also employed: \(\textstyle{\mathit{HD(\mathit{GT}, \mathit{P})}=\max(\mathit{h(\mathit{GT}, \mathit{P})}, \mathit{h(\mathit{P}, \mathit{GT})})}\) with \(\textstyle{\mathit{h(\mathit{A}, \mathit{B})}= \max_{\mathit{a} \in \mathit{A}}\min_{\mathit{b} \in \mathit{B}}\,\|\mathit{a}-\mathit{b}\|}\).

\section{Results and discussion}

\begin{figure}[t]
    \centering
    \includegraphics[height=6.6cm]{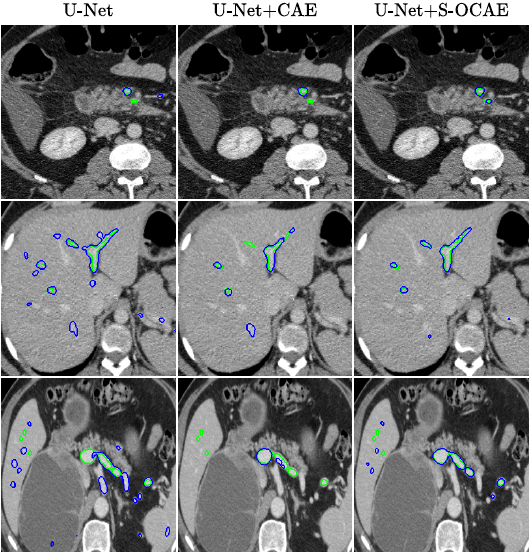}
    \caption{Liver vessel segmentation results. Ground truth and predicted contours are respectively in green and blue.}
    \label{fig:qualres_ircad} \vspace{-0.2cm}
\end{figure}

\begin{figure}[t]
    \centering
    \includegraphics[height=6.6cm]{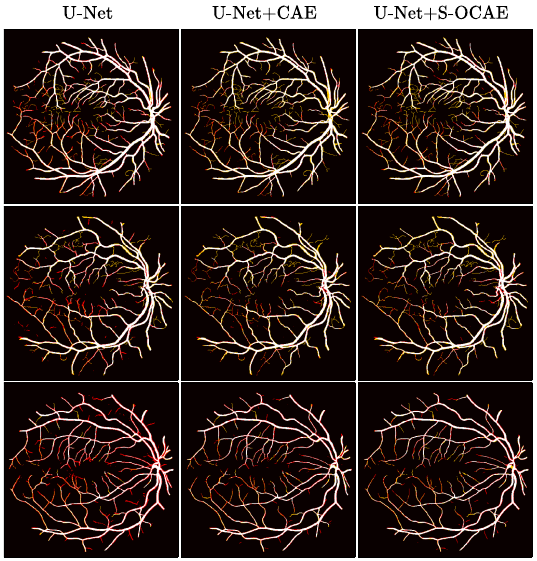}
    \caption{Retinal vessel segmentation results. True positives, false positives and false negatives are in white, red and yellow.}
    \label{fig:qualres_drive} \vspace{-0.25cm}
\end{figure}

U-Net \cite{ronneberger2015unet} using VGG13 with BN pre-trained on ImageNet as encoder \cite{conze2021abdominal} was compared with the same U-Net architecture complemented with shape priors from either a standard CAE (U-Net+CAE) \cite{oktay2017anatomically} or the proposed S-OCAE (U-Net+S-OCAE). Both auto-encoders were trained from scratch. 

Quantitative results (Tab.\ref{Tab:table}) indicate that adding shape priors to U-Net significantly improves delineation results in all assessment metrics. Moreover, whatever the dataset, the proposed method outperforms U-Net+CAE in \texttt{DSC}, \texttt{AVD}, \texttt{ASSD}, and \texttt{HD} except for 3D-IRCADb where U-Net+CAE reaches a slightly better \texttt{DSC}. The main improvement from U-Net+CAE to U-Net+S-OCAE is observed at the level of the \texttt{HD} metric which undergoes a rise of $3.08\%$ for DRIVE and $4.03\%$ for 3D-IRCADb. The attenuation of  the largest segmentation errors suggests that vessel contours provided by U-Net+S-OCAE are more suitable with respect to clinical requirements. Results without and with CB reveal the positive impact of making communicate under and overcomplete features. In addition, vessel delineations displayed in Fig.\ref{fig:qualres_ircad} and \ref{fig:qualres_drive} highlight a better ability of U-Net+S-OCAE to extract small structures against other methods, while providing less false positives. These results tend to indicate that relying on a semi-overcomplete embedding to produce shape priors leverages a robust representation of the vascular tree geometry, despite the large range of various multi-scale structures visible in both DRIVE and 3D-IRCADb datasets.

\section{Conclusion}
\label{sec:conclusion}

In this work\footnotemark[3], we proposed a fully automatic segmentation pipeline which incorporates anatomical shape embedding as priors from a newly designed Semi-Overcomplete Convolutional Auto-Encoder (S-OCAE). Resulting shape priors demonstrated their usefulness to drive the segmentation model to constraint the largest delineation errors, hence providing a robust blood vessel extraction tool. In future works, the combination of both geometric and topological constraints will deserve further investigation to improve the ability of deep models to provide more realistic vessel contours.

\footnotetext[3]{This research study was conducted retrospectively using human subject data made available in open access  \cite{ staal2004ridge,soler20103d}.}

\bibliographystyle{IEEEbib}
\bibliography{references}

\end{document}